\documentstyle[preprint,eqsecnum,aps]{revtex}
\begin{document}
\draft
\preprint{UCSBTH-96-13, hep-th/9606113}
\title{Counting  States  of Black Strings with Traveling Waves II}
\author{Gary T. Horowitz\cite{gary} and Donald Marolf\cite{don}}
\address{Physics Department, University of California,
Santa Barbara, California 93106} \date{June, 1996}
\maketitle

\begin{abstract}
We extend our recent analysis of the entropy of extremal black strings with
traveling waves. We previously considered waves carrying
linear momentum on black strings in six dimensions. Here we study waves
carrying angular momentum on these strings, and also waves
carrying linear momentum on black strings in five dimensions.
In both cases, we show that the horizon remains homogeneous and
compute its area. We also count the number of BPS
states at weak string coupling
with the same distribution of linear and angular momentum,
and find complete agreement with the black string entropy.  

\end{abstract}

\newcommand{\f}{\dot f}
\newcommand{\k}{\kappa^2}
\newcommand{\al}{\alpha}
\newcommand{\p}{\partial}
\newcommand{\V}{{\sf V}}
\newcommand{\be}{\begin{equation}}
\newcommand{\ee}{\end{equation}}

\vfil
\eject

\baselineskip = 15pt
\section{Introduction}

In a recent paper \cite{homa}, we studied a class of supersymmetric
solutions to string theory which contain regular event horizons
and depend on arbitrary functions. These solutions describe extremal
black strings
with traveling waves and have an inhomogeneous distribution of momentum
along the string. (Solutions of this type were first discussed
in \cite{lawi}.) It was shown that for each traveling wave,
the Bekenstein-Hawking entropy 
agreed precisely with the number of BPS states at
weak string coupling having the same momentum distribution as the black string.
This extended recent work [3 - 11] in which the microstates corresponding
to black hole entropy were identified. 
These earlier investigations reproduced  the gravitational entropy of certain
black holes (or translationally invariant black strings) by counting
the number of bound states of D-branes \cite{pol,pcj} with fixed total momentum.
We found that this agreement
extends to the case where an (essentially) arbitrary momentum distribution
is fixed and the corresponding black string is not translationally invariant.

In \cite{homa}, we considered six dimensional black strings with a constant
internal four dimensional space (so that the total spacetime has ten
dimensions). We studied two different types
of  traveling waves; one carrying momentum along the string, and
the other carrying momentum both along the string and
in the internal four dimensional space.
Here, we extend this work in two directions. In section II we
add waves carrying angular momentum to the six dimensional black string.
We compute the horizon area and compare the Bekenstein-Hawking entropy
with the number of D-brane states at weak coupling with a 
given angular momentum distribution. Once again we find complete agreement.
In section III, we consider five dimensional black strings (which yield four
dimensional black holes upon dimensional reduction).  To
generalize the treatment,
the size of the internal five torus is allowed to vary in
spacetime. We study waves 
carrying linear\footnote{ 
Rotating black holes in four dimensions 
are not supersymmetric and five dimensional black strings do not
support traveling waves with angular momentum.} momentum in all
spacelike directions
and examine the solutions near the horizon. We then
compare the horizon area and the
number of D-brane states with the given momentum
distributions.
As before, the Bekenstein-Hawking entropy of the black string agrees 
with the counting of states.

It was also shown in \cite{homa} that the traveling waves do not affect the
local geometry of the event horizon; it remains a homogeneous surface.
Physically, this is because the waves become purely outgoing near the horizon.
We will see that the same remains true for all the waves studied here.

\section{Angular Momentum Waves for 6D Black Strings}
\label{rot}

As in \cite{homa}, we consider 
six dimensional black string solutions to type IIB string theory
compactified on a four torus with volume $\V $.  We assume one 
additional spatial direction is compactified to form a circle of length $L$
and choose $ L \gg \V^{1/4}$ so that the solutions resemble strings
in six dimensions. We are interested in solutions with
nonzero electric and magnetic charges
associated with the RR three-form; in the limit of weak string coupling,
these charges are carried by D-onebranes and D-fivebranes.

Solutions with these charges
have a regular event horizon even in the extremal limit.
Furthermore, such extremal black strings have a null Killing field 
${\partial /{\partial v}}$ so that one can use the
observations of \cite{GV,Garf} to add
traveling waves. Several types of
waves carrying linear momentum were considered in \cite{homa}.
We now study a different class of 
waves traveling along the string; these waves will carry
angular momentum. 

\subsection{Traveling Waves Carrying Angular Momentum}

In \cite{homa}, we investigated the metric
\begin{equation}
\label{old}
ds^2 = \left(1 + {{r_0^2} \over {r^2}}\right)^{-1} \left[-dudv +
 \left( {{p(u) } \over {r^2}} - 2 \ddot{f}_i(u) y^i 
- 2 \ddot{h}_i(u) x^i \right) du^2 \right]
+\left(1 + {{r_0^2} \over {r^2}}\right) dx_i dx^i +dy_i dy^i 
\end{equation}
which describes a black string carrying a `longitudinal 
wave' $p(u)$, an `internal wave' $f_i(u)$, and an `external wave'
$h_i(u)$.  Each of these waves carry  momentum in a different
direction \cite{cmp,DGHW}.  In the metric (\ref{old}), dots denote $d/du$,
the coordinates $y^i$ label points on the $T^4$,
the $x^i$ label points in the four dimensional
asymptotically flat space, and $r^2 = x_i x^i$.  In addition, 
$u=t-z, \
v=t+z$ where $z$ is a
coordinate on the $S^1$.
For this solution, the dilaton is constant and
the (integer normalized) charges associated with the
RR three-form take the values
\begin{equation}
Q_1 = {Vr_0^2\over g}, \qquad Q_5 = {r_0^2\over g}
\end{equation}
where $g$ denotes the string coupling. 

We may now add angular momentum to this black
string as described in \cite{tse1}. To preserve supersymmetry, one needs
equal amounts of angular momentum in two orthogonal planes \cite{bmpv}.
Although \cite{tse1} considered
only uniformly rotating
strings,
the angular momentum density
may again be taken to be a function of $u$ without otherwise changing
the metric or matter fields \cite{cvts}.
This is directly analogous to the situation
for longitudinal momentum. 
Writing the metric on the three sphere in the form 
$d\Omega^2_3 = d\theta^2 + \sin^2 \theta d \varphi^2
+ \cos^2 \theta d \psi^2$, we obtain 
\begin{eqnarray}
\label{rotwave}
ds^2 &=& \left(1 + {{r_0^2} \over {r^2}}\right)^{-1} \left[-dudv +
 {{p(u) } \over {r^2}}  du^2 
+  {2{\gamma(u)} \over {r^2}}
(\sin^2 \theta d \varphi - \cos^2 \theta d \psi ) du \right] \cr
&+&\left(1 + {{r_0^2} \over {r^2}}\right) (dr^2 + r^2 
d\Omega^2_3) +dy_i dy^i, 
\end{eqnarray}
which describes a black string with angular momentum density
$\gamma(u)/\kappa^2$. The constant
$\kappa^2$ is given by
\be\label{defkap}
\k \equiv {4G_{10} \over  \pi\V} = {2\pi g^2 \over  V},
\ee
where we have used the fact that the ten dimensional Newton's constant is
related to the string coupling by $G_{10} = 8\pi^6 g^2$ in units with
$\alpha' =1$ and set $\V = (2\pi)^4 V$. This black string has total
angular momentum
\be
J_\varphi = - J_\psi = \kappa^{-2} \int_0^L \gamma(u)\ du
\ee
and longitudinal momentum
\be
P= \kappa^{-2} \int_0^L p(u) \ du.
\ee
For simplicity, we have set the internal wave $f_i(u)$ and
the external wave $h_i(u)$ to zero in (\ref{rotwave}). These waves could have
been retained without altering the conclusions below.

The longitudinal wave $p(u)$, on the other hand, 
cannot be set to zero in the presence of angular momentum.
We will see that, in order for the horizon to have a finite area, 
the longitudinal wave $p(u)$ must be at least $\gamma^2(u)/r_0^4$
at each point along the string.  Of course, both $p(u)$
and $\gamma(u)$ must be periodic with period $L$.

It turns out that, near the horizon, this metric effectively
coincides with a metric studied in \cite{homa}.  As a result, after
making the correspondence clear, we may simply read off the desired
features of the horizon.  This is done by replacing $\phi, \psi$
with the new coordinates
\begin{eqnarray}
\label{angles}
\tilde{\varphi} &=& \varphi + {{\beta(u)} \over{(r^2 + r^2_0)^{2}}}, \cr
\tilde{\psi} &=& \psi - {{\beta(u)} \over {(r^2 + r^2_0)^{2}}}, \cr
{\rm with} \ \beta(u) &=& \int^u \gamma(u')\  du'.
\end{eqnarray}
The metric then takes the somewhat complicated form
\begin{eqnarray}
\label{long}
ds^2 &=& \left(1 + {{r_0^2} \over {r^2}}\right)^{-1} \left[-dudv +
\left({{p(u) } \over {r^2}} - {{\gamma^2(u) } \over
{r_0^4 r^2}}\right) du^2\right] \cr
&+& \left(1 + {{r_0^2} \over {r^2}}\right) \left(
dr^2 
+ r^2 [d \theta^2
+ \sin^2 \theta d \tilde{\varphi}^2 + \cos^2 \theta d \tilde{\psi}^2]
\right)
+dy_i dy^i  \cr
&+& 
{{16 r^2\beta^2
} \over {(r^2 + r^2_0)^5} } dr^2
+ {8 \beta
r \over {(r^2 + r_0^2)^2}}dr (\sin^2 \theta d\tilde{\varphi}
- \cos^2 \theta d\tilde{\psi})  \cr
&+& {{r^2 \gamma^2} \over {r_0^4(r^2 + r_0^2)^3}} (r^2 + 2 r_0^2) du^2.
\end{eqnarray}
Note, however, that the first two lines of (\ref{long}) give just the
metric (\ref{old})
for a black string with the longitudinal wave $p - \gamma^2/r_0^4$
and no angular momentum.
The last two lines may be considered as correction terms;  they
are all of sub-leading order near the horizon.
In fact, the terms on the third line of (\ref{long})
vanish on the horizon and require no further work.  Because the
horizon lies at $u = \infty$, the
$du^2$  term (on the last line above) does not vanish
on the horizon; however,
this term turns out to be 
of the same form as the subleading order terms discussed in
the appendix of \cite{homa}.  As a result, the techniques used there
apply to this case as well and show both that the metric is at least
$C^0$ at the horizon and that the horizon is locally a homogeneous surface.
Again, the waves do not affect the local horizon geometry.

To compute the area of the horizon, we must study both 
its local {\it and} its global structure.  From (\ref{angles}),
we may deduce the effects of our coordinate transformation 
on the global identifications.  Note that the new angles
$(\tilde{\varphi},\tilde{\psi})$ 
remain periodic with period
$2\pi$ on the three-sphere, but that 
under the identification $z \rightarrow z - L$
the new angular coordinates acquire a shift:
$(\tilde{\varphi},\tilde{\psi}) \rightarrow (\tilde{\varphi} + 
(r^2 + r^2_0)^{-2} \int_0^L
\gamma du, \ \tilde{\psi} - (r^2 + r^2_0)^{-2} \int_0^L \gamma du)$. 
This change in global structure
is analogous to changing the modular parameter of a
torus; the global structure is different, but the volume remains
unchanged.  In this case, the new identifications imply
that a translation along the horizon
is accompanied by a rotation of the three-sphere in the two orthogonal
planes.  It follows that the
horizon area is the same as for
a longitudinal traveling wave with profile 
\be
\tilde{p}(u) =p(u) - {\gamma^2(u)\over r_0^4}. 
\ee

As explained in \cite{homa}, to write this area in a 
simple form we must introduce a function $\sigma$, periodic in
$u$, which is related to $\tilde{p}(u)$ by $\sigma^2 + \dot{\sigma} = 
r_0^{-4} \tilde{p}$.   The horizon area is then
\be
A = 2\pi^2 r_0^4 {\sf V} \int_0^L \sigma(u)\ du.
\ee
When $\tilde{p}$ satisfies the `slowly varying condition'
\begin{equation}
\label{slow5}
\tilde{p}^{3/2} \gg r_0^2 |\dot{\tilde{p}}|,
\end{equation}
we have $\sigma = \sqrt{\tilde{p}/r_0^4}$ and the area takes the form
$A = 2\pi^2 r_0^2 {\sf V} \int_0^L \sqrt{\tilde{p}} du $.
The corresponding Bekenstein-Hawking
entropy is then
\begin{equation}
\label{simple5}
S_{BH} = {A \over {4G_{10}}} = 
\sqrt{2\pi Q_1Q_5} \int_0^L \sqrt{\tilde{p}(u)/\kappa^2}\  du.
\end{equation}
In the special case where $p$ and $\gamma$ are constant, the total longitudinal
momentum is $P = pL/\k$ and the total angular momentum is $J = \gamma L/\k$.
Setting $P= 2\pi N/L$, we obtain $S_{BH} = 2\pi \sqrt{Q_1 Q_5 N - J^2}$ 
as expected \cite{bmpv,tse1}.

\subsection{Counting BPS States}

We now show that the exponential of the entropy (\ref{simple5}) yields the
number of BPS states at weak string coupling with the same distribution
of angular momentum and longitudinal momentum as the black string.
Recall that at weak coupling the black string corresponds to a collection
of D-fivebranes and D-onebranes. The low energy excitations are described by
a supersymmetric sigma model in $1+1$ dimensions (where the
spatial direction corresponds to our `string' direction $z$) 
containing  $4 Q_1 Q_5$
bosonic fields and an equal number of fermionic fields.

It was shown in \cite{bmpv} that the number of 
D-brane configurations with total momentum $P$ and angular momentum $J$
agrees with the Bekenstein-Hawking entropy of an extreme five dimensional
black hole, which is equivalent to a homogeneous black string, i.e.
(\ref{rotwave}) with $p=P\k/L,\ \gamma = J\k/L$.
Our strategy will be to consider the
string to be made up of many short homogeneous segments
and then to 
apply the results of \cite{bmpv} to each one in turn.  The key point is
thus to show that the fields on different segments may be treated
independently.  The argument is identical to the one
given in \cite{homa}, so we
will only sketch the derivation below.

Recall that one cannot fix the
{\it exact} value $j(u)$ of a current in a 1+1 quantum field theory.
The reason is simply that $j(u)$ and $j(u')$ do not commute; 
for example, when $j$ is the momentum density, its Fourier
modes satisfy the Virasoro algebra.
We therefore
take a `mesoscopic' viewpoint for our discussion.  That is, we imagine
that we use an apparatus which can resolve the system only down to a
`mesoscopic' length scale $l$ which is much larger than the
`microscopic' length scale (discussed below) on which quantum effects
are relevant.  We will therefore divide the spacetime into
$ L/l$ intervals $\Delta_a$ ($a \in \{1,...,L/l\}$) of length
$l \ll L$.  If our instruments find a momentum distribution $
 p(u)/\k$ and an angular momentum distribution $
\gamma(u)/\k$,
this simply means that the interval $\Delta_a$ contains a momentum
$P_a =\kappa^{-2} \int_{\Delta_a}p(u) du$ and an angular momentum
$J_a = \kappa^{-2} \int_{\Delta_a} \gamma(u) du$;
we cannot resolve $p$ and $\gamma$
on
smaller scales.  Of course, it would be meaningless for us to
assign a distribution $(p,\gamma)$ which has structure on scales of
size $l$ or smaller.  As a result, $l$ should be much smaller
than $p/|\dot{p}|$ and $\gamma/|\dot{\gamma}|$ (and
therefore $\tilde{p}/|\dot{\tilde{p}}|$), the `macroscopic' length
scales set by the
variation of the wave profile $(p,\gamma)$.

We shall take the idea that $l$ is much larger than any microscopic
length scale to mean that the `level numbers'
$P_a l$  and
$(P_a l - J_a^2 \kappa^2 r_0^{-4})$ are both large
($\gg Q_1 Q_5$).  These 
conditions imply that the wavelength
of a typical excited mode with momentum $\tilde p l/\k$ is much less than $l$,
and they are just the
conditions imposed in \cite{bmpv} to enable a counting of states
on a string of length $l$.  Thus we must choose $l$ to satisfy
$\tilde{p}/\dot{\tilde{p}}
\gg l \gg \sqrt{Q_1 Q_5\kappa^2 / \tilde{p}}$. Such an $l$ can exist only when
\begin{equation}
\tilde{p}^{3/2} \gg |\dot{\tilde{p}}| \sqrt{Q_1Q_5 \kappa^2}
= \sqrt{2 \pi} r_0^2 |\dot{\tilde{p}}|,
\end{equation}
which is equivalent to the slowly varying condition 
(\ref{slow5}). 

Under this condition, the arguments of \cite{homa}
show that the entropy is carried by modes of sufficiently high 
frequency that each interval of length $l$ may
be treated as a separate sigma model. 
The counting of BPS states then reduces to considering 
states of longitudinal momentum $(P_a - {{J_a^2 \kappa^2}
\over  {lr_0^{4}}})$
and {\it no} angular momentum; that is, the angular
momentum $J$ affects the entropy only by reducing the momentum
to be distributed among the entropy carrying modes.  Each
segment then carries an entropy of $S_a = \sqrt{2 \pi Q_1Q_5 }
\sqrt{P_a l - {{J_a^2 \kappa^2} \over {r_0^4}}}$ and
the total entropy is
\begin{equation}
S = \sqrt{2 \pi Q_1 Q_5} \int_0^L \sqrt{\tilde{p}/\kappa^2}\ du.
\end{equation}
Thus the number of BPS states agrees with
the Bekenstein-Hawking entropy (\ref{simple5}) of a black string with the
corresponding distribution of momentum and angular momentum.

\section{Traveling Waves for 5D black strings}
\label{four}

We now turn to solutions of type IIA string theory which describe
black strings in five dimensions. These solutions are related to 
{\it four} dimensional black holes. In close analogy with the 6D black
strings, one can add traveling waves to the extremal 5D black strings.
We will show that the Bekenstein-Hawking entropy of these solutions
again corresponds to the number of BPS 
states at weak string coupling with the same distribution
of energy and momentum. This extends the results of \cite{mast,jkm}
where this correspondence was shown for the extremal
black strings without traveling
waves. In addition, we expand the discussion to
allow certain moduli of the internal torus
to vary across the spacetime.  

We will consider solutions to type IIA string theory carrying magnetic 
charge with respect to the RR two-form, electric charge with respect
to the RR four-form, and magnetic charge with respect to the NS-NS
three-form. At weak coupling, these charges are carried by a D-sixbrane,
D-twobrane, and solitonic fivebrane, respectively\footnote{One can
also form five dimensional black strings or four dimensional black holes
with other choices of charges \cite{jkm,klts,bala}. This choice was first
used in \cite{mast}. We follow the conventions of \cite{hlm}.}. We take  six
dimensions to be compactified to form a torus and assume translational
symmetry in five of these dimensions. The sixth ($z$) direction
will have a length $L$ much
longer than the others, and will be the direction in which the waves propagate.
Hence these solutions describe black strings with traveling waves
in five dimensions. Four of the toroidal directions (labeled by
$y^a ,\  a = 1,2,3,4$) form a torus of 
volume ${\sf V} = (2\pi)^4 V$,
and one ($w$) forms
a circle of length $\tilde{L}$.
The other 
four dimensions ($t,x^i,\ i=1,2,3$) will be asymptotically
flat.  

\subsection{Classical Black String Solutions}
\label{cs}

The solution of type IIA string theory describing five dimensional
black strings with the above charges 
was found in \cite{tsyhar}. It is characterized by
three harmonic functions 
\be
H_2(r) = 1+ {r_2 \over r}, \qquad H_5(r) = 1+{r_5 \over r}, 
\qquad H_6(r) =1+{r_6\over r},
\ee
where $r^2 = x_ix^i$.
The constants $r_2,\ r_5,\ r_6$ are related to the integer charges by
\be\label{chargefd}
Q_2 = {2r_2 V\over g}, \qquad Q_5 = {r_5 \tilde L\over \pi},
\qquad Q_6 = {2r_6\over g}.
\ee
Introducing the null coordinates $u = t - z$ and $v = t + z$
and using the flat metric $\delta_{ij}$ and $\delta_{ab}$ to
raise and lower the indices $i \in \{1,2,3\}$ and $a \in \{1,2,3,4\}$,  
the Einstein metric for the black string takes the form
\begin{eqnarray}
\label{harm}
ds^2 &=& H_2^{3/8} H_5^{6/8} H_6^{7/8}\Big[-H_2^{-1} H_5^{-1} H_6^{-1}
du dv \cr &+& H_5^{-1}
H_6^{-1}dy_a dy^a
+H_2^{-1} H_6^{-1} dw^2 + dx_idx^i\Big]
\end{eqnarray}
and the dilaton is $e^{2\phi} = H_2^{1/2} H_5 H_6^{-3/2}$.
The horizon is at $r=0$, $u = \infty$.
Note that $\phi$ approaches a constant both at infinity and
at the horizon.
When  all three harmonic functions are equal, $H_2 = H_5 = H_6\equiv H$,
the dilaton
vanishes and the metric reduces to
\be
ds^2 = -H^{-1} du dv  + dy_a dy^a + dw^2 + H^2 dx_idx^i
\ee
which is just the product of a five torus and the
extremal black string solution of the five dimensional
Einstein-Maxwell theory \cite{ght}.

Since the solutions (\ref{harm}) all possess the null Killing vector
field ${\partial} / {\partial v}$, we may again add traveling
waves using the methods of \cite{GV,Garf}.
The result is a metric of the form
\begin{eqnarray}
\label{4waves}
ds^2 &=& H_2^{3/8} H_5^{6/8} H_6^{7/8}\Big[H_2^{-1} H_5^{-1} H_6^{-1}
du \left[-dv + K(u,x,y) du \right] \cr &+& H_5^{-1}
H_6^{-1}dy_a dy^a
+H_2^{-1} H_6^{-1} dw^2  + dx_idx^i\Big]
\end{eqnarray}
where $K$ 
satisfies
\begin{equation}
\label{K}
(\partial_i \partial^i + \partial_a \partial^a +
{\partial}_w
{\partial}_w
) K = 0.
\end{equation}
That is, $K$ is harmonic in the toroidal ($w, y^a$) and 
asymptotically flat ($x^i$) coordinates, but has arbitrary $u$ dependence.
As a result, 
$K$ contains free functions that describe traveling waves
along the `string direction' ($z$).  Since the surface $r=0$
is a coordinate singularity in (\ref{harm}), we  only require (\ref{K})
to hold for $r\ne 0$. Nonetheless, the
horizon will be a regular surface for all of the metrics we consider.
For the moment, we will
ignore the details of the compactifications; they will be 
discussed below.

We wish to consider only waves 
that are in some way `anchored' to the black string, i.e.,
waves that either become pure gauge or have unphysical singularities
when the black string is removed. Such waves were first discussed in
\cite{cmp,DGHW} in connection with a fundamental string and later in \cite{homa}
for a six dimensional black string. In the present context, the waves are 
given by
\begin{equation}
\label{waves}
K = {{p(u)} \over {r} }
-2 \ddot{f}_a(u) y^a - 2 \ddot{b}(u)w - 2 \ddot{h}_i(u) x^i.
\end{equation}
As before,  the $p$ term  represents `longitudinal
waves,' the $f_a$ and $b$ terms  represent `internal waves', and
the $h_i$ term represents `external waves.'
We will see that
this terminology corresponds to the various directions in which
the waves carry momentum.  Since the internal and external waves are
clearly negligible compared to the longitudinal wave 
near the horizon ($r=0$), one might expect that they do not contribute to
the horizon area. We will see that this is indeed the case.

With the waves (\ref{waves}), the metric (\ref{4waves}) is neither
asymptotically flat nor translationally invariant in the toroidal
directions.  Both  difficulties can be resolved by transforming
to coordinates $(u,v',w',x',y')$ which are related to those above
through
\begin{eqnarray}
v' &=& v + 2 \dot{f}_a y^a  + 2 \dot{b} w + 2 \dot{h}_i x^i
+ \int^u (\dot{f}^2 + \dot{b}^2 + \dot{h}^2) du, \cr
w' &=& w + b, \cr
x'{}^i &=& x^i + h^i, \cr
y'{}^a &=& y^a + f^a, 
\end{eqnarray}
where $\dot{f}^2 = \dot{f}_a \dot{f}^a$, $\dot{h}^2 = \dot{h}_i
\dot{h}^i$.  
The metric then takes
the form
\begin{eqnarray}
\label{af}
ds^2 &=& H_2^{3/8} H_5^{6/8} H_6^{7/8}\Bigg[H_2^{-1} H_5^{-1} H_6^{-1} du
\Bigg(-dv' + \bigg[ {{p+r_2 \dot{f}^2 + r_5 \dot{b}^2}
\over r} + (H_2H_5H_6 -1) \dot{h}^2 \bigg] du \cr
&-&  {2{r_2} \over {r}} \dot{f}_a dy'{}^a -  {2{r_5} \over r} \dot{b} dw'
- 2 (H_2H_5H_6 -1 ) \dot{h}_i dx'{}^i
 \Bigg) \cr &+& H_5^{-1}
H_6^{-1}dy'_a dy'{}^a
+H_2^{-1} H_6^{-1} dw' dw' + dx'_idx'{}^i\Bigg].
\end{eqnarray}
It is in terms of these coordinates that we make the periodic
identifications which compactify the spacetime. Setting $z'=(v'-u)/2$, the
large $S^1$ is defined by the identification $z' \rightarrow z' - L$, 
or $(u,v',w',x',y') \rightarrow (u + L, v' -L,w', x',y')$, while the
small five-torus is defined by the identifications $w' \rightarrow w' + 
\tilde L$, and $y' \rightarrow y' + a_I$
for an appropriate set of four vectors $a_I$.
Clearly, $p(u)$, $\dot{f}(u)$, $\dot{b}(u)$, 
and $\dot{h}(u)$ must be periodic in $u$.  In addition, we will
take $f(u)$, $b(u)$, and $h(u)$ to be periodic themselves.
In the ten-dimensional
space before compactification, this amounts to considering only
black strings with no net momentum in the $w'$, $x'$, and $y'$ directions.

The asymptotic charges can be read directly from the metric (\ref{af}).
As in \cite{homa}, the momentum is not 
distributed uniformly along the string, and we will match this
momentum profile to a configuration of D-branes at weak coupling.
Defining\footnote{Although $\k$ will play the same role in this section
as it did in the previous one, its precise definition is different.
In section II, $\k$ was related to the six dimensional Newton's constant;
here it is related to Newton's constant in five dimensions.}
\be\label{kappafd}
\kappa^2 \equiv {4 G_{10}\over {\sf V} \tilde L} =
{2\pi^2 g^2 \over V \tilde L},
\ee
the black string with traveling waves (\ref{af}) has ADM momentum 
\begin{eqnarray}
\label{momen}
P_{z'} &=& \kappa^{-2} \int_0^L du
\left[ p + r_2 \dot{f}^2 + r_5 \dot{b}^2
+ (r_2 + r_5 + r_6)\dot{h}^2  \right]\cr
P_a &=& \kappa^{-2} r_2 \int_0^L du \ \dot{f}_a  \cr
P_{w'} &=& \kappa^{-2} r_5 \int_0^L du \ \dot{b} \cr
P_i &=& \kappa^{-2} (r_2 + r_5 + r_6)\int_0^L du\ \dot{h}_i \ ,
\end{eqnarray}
and ADM energy
\be
E = \kappa^{-2} ( r_2 + r_5 + r_6) L + P_{z'} \ .
\ee
It is clear from (\ref{momen}) that the black string is composed of
three different constituents: 
Oscillations of the same amplitude in different directions
result in different amounts of momentum. This is consistent with the
weak coupling description in terms of branes wrapped around different
directions,
but is {\it not} one would expect from, say, 
a large rubber band.  The spacetime metric thus records the fact that
any `source' must have several components.

\subsection{The Event Horizon}

We would like to show 
that $r=0$, $u = \infty$ is a horizon in the spacetime (\ref{4waves})
and to compute the horizon area. 
The calculations are structurally identical to
those performed in \cite{homa} but are slightly more complicated
due to the presence of the different harmonic functions $H_2$, $H_5$,
and $H_6$.   We will not present the full details here, but
the reader may reconstruct them by copying the steps 
described in the appendix of \cite{homa}.  Such
techniques suffice to show that the
horizon is at least $C^0$ and its area may
be computed using only the leading order behavior 
of the metric  (\ref{4waves}) near $r=0$. Setting $r= 4r_2 r_5 r_6/R^2$,
this leading order metric takes the form
\begin{equation}
ds^2 = (r_2^3 r_5^6 r_6^7)^{1/8} \left[4\left( R^{-2}[-du dv + dR^2]
+ {p(u) \over 4 r_2r_5 r_6} du^2\right) + d\Omega_2^2
+ {{dy_a dy^a} \over{r_5 r_6}} +
{{dw^2} \over {r_2 r_6}} \right].
\end{equation}
Using the results in \cite{homa} one finds that this spacetime has a homogeneous
horizon with area
\be
A = 8 \pi r_2 r_5 r_6 \tilde{L} {\sf V} \int_0^L \sigma(u)\ du
\ee
where $\sigma$ is a periodic function of $u$ satisfying
$\sigma^2 + \dot{\sigma} =  p/(4r_2 r_5 r_6)$.
In analogy with the six dimensional
case, when
\begin{equation}
\label{slow4}
p^3 \gg (r_2r_5r_6) \dot{p}^2,
\end{equation}
the area becomes $A = 4\pi \sqrt{r_2r_5r_6}
\tilde{L} {\sf V} \int_0^L \sqrt {p(u)} \ du$
and the Bekenstein-Hawking entropy may be
written
\begin{equation}
\label{BH4}
S_{BH} = \sqrt{2 \pi  Q_2 Q_5 Q_6} \int_0^L \sqrt{p(u)/\kappa^2}\ du.
\end{equation}

\subsection{Counting BPS States}

Consider the weak coupling limit of the IIA string theory with six dimensions
compactified to form a torus as above; one circle has length $L$ much larger
than the rest, another
has length $\tilde L$ and the remaining four have volume $\V = (2\pi)^4 V$.
The black strings (\ref{harm}) correspond to $Q_2$
D-twobranes wrapped around $L$ and $\tilde L$, $Q_5$ solitonic fivebranes
wrapped around $L$ and $\V$, and $Q_6$ D-sixbranes wrapped around the entire
six torus. Each brane contributes an effective string tension in the $L$ 
direction given by
\be
T_2 = {\tilde L \over 4\pi^2 g}, \qquad T_5 = {V\over 2\pi g^2},
\qquad T_6 = {V\tilde L\over 4\pi^2 g}.
\ee
It then follows from (\ref{chargefd}) and (\ref{kappafd})  that
\be\label{qt}
{r_2\over \kappa^2} = Q_2 T_2, \qquad {r_5\over \kappa^2} = Q_5 T_5, \qquad
{r_6\over \kappa^2} = Q_6 T_6.
\ee

 We begin by setting
$b = h_i = 0$, and proceed as in section III of \cite{homa}.
We again adopt a mesoscopic viewpoint and divide the string into
a number of small intervals.  We will consider only the case
$Q_6 = 1$, in which the
degrees of freedom that contribute to the entropy
correspond to oscillations of the twobranes in the sixbranes. 
There are actually $4Q_2Q_5$ (bosonic) degrees of freedom of this type
since the fivebranes `cut' each twobrane into $Q_5$ different
pieces \cite{mast}, each with an effective average tension of $T_2/Q_5$
in the string direction.
We interpret the {\it field} momentum of these bosonic fields as
carrying {\it spacetime} momentum in the internal $y^a$ directions since
both generate translations of the twobranes.  
Recall that a field $\chi$ with tension $T$ has momentum density $T\dot{\chi}$,
 and the black string has momentum density $r_2 \dot{f}_a/\k$ in the
 $y^a$ direction (\ref{momen}). Thus
the condition for the $Q_2 Q_5$ fields $\chi_A$
associated with fluctuations in the $y_1$
direction to have the
 same momentum as the black string is
\begin{equation}
\label{constraint}
{T_2 \over Q_5} \sum_{A=1}^{Q_2Q_5} \dot{\chi}_A  = 
{r_2{\dot{f}_1} \over
{\kappa^2}} = Q_2 T_2 \dot{f}_1.
\end{equation}
It follows that $f_1$ is just the average fluctuation of the D-branes.
We impose similar conditions for the other three internal directions
($y_2, y_3, y_4$) transverse to the twobranes.

The counting may now be performed just as in 
\cite{homa}.  Applying our `mesoscopic' viewpoint, we
divide the string into segments of length $l$ and restrict only the
average values of the oscillations $\chi_A$ to satisfy (\ref{constraint})
on each segment.
Thus the internal momentum  is carried by the zero mode in each
segment. An
internal momentum distribution $Q_2 T_2\dot{f}^a$ among $Q_2Q_5$
twobranes of tension $T_2/Q_5$ requires a longitudinal momentum of at least
$Q_2 T_2\dot{f}^2 $. The remaining longitudinal momentum is just $p/\k$,
and the entropy arises from distributing this momentum arbitrarily.
When each interval is highly excited, this gives the result
\begin{equation}
\label{4BPSS}
S = \sqrt{2 \pi Q_2 Q_5} \int_0^L \sqrt{p(u)/\kappa^2}\ du.
\end{equation}
As usual, the intervals can be highly excited only when
$p^3 \gg (r_2 r_5 r_6) \dot{p}^2$ so that the slowly varying condition
(\ref{slow4}) holds.  Since $Q_6=1$,
(\ref{4BPSS}) agrees with the Bekenstein-Hawking
entropy (\ref{BH4}) of the black string.

Now  consider the waves  $b(u)$ and $h_i(u)$. 
In general, the oscillation $\chi$ of a string with tension $T$ has momentum
$T \dot \chi$ in the transverse direction and $T \dot{\chi}^2$
in the longitudinal direction. It thus follows from (\ref{momen}) and (\ref{qt})
that the wave $b(u)$ is
naturally interpreted as corresponding to macroscopic oscillations
of the fivebranes, and the wave $h_i(u)$ is interpreted as coordinated
oscillations of all the branes together in the macroscopic directions. 
The contributions of $b(u)$ and $h_i(u)$ to both the longitudinal and transverse
momenta can be accounted for in this way. 
At weak string coupling, there are a small number of fields which
describe these oscillations. They act just like the fields $\chi$
described above: By requiring that their field momenta agree
with the transverse momentum of the black string,  we `use up' part of the 
longitudinal momentum. The remaining longitudinal momentum is just
$p(u)/\k$, independent of  $b(u)$ and $h_i(u)$. Since the number of
such fields is small compared to $Q_2 Q_5$, 
they do not change the number of degrees of freedom
which can carry the longitudinal momentum to leading order. Thus the entropy is
again given by (\ref{4BPSS}) in agreement with the Bekenstein-Hawking
entropy of the black string.

\section{Discussion}

We have  considered two new families of solutions 
containing black strings with traveling waves. One describes six dimensional
black strings with waves carrying angular momentum, and the other
describes five dimensional black strings with  waves carrying linear momentum
in various directions.  We computed the horizon area,
and counted the number of
BPS states at weak string coupling with the same distribution
of momentum and angular momentum. In each case, the Bekenstein-Hawking
entropy agreed with the number of microscopic string states.
Combined with the results of \cite{homa}, this leads to the following
conclusion:
{\it In every case where the entropy
of an extreme
black hole has been understood in terms of string states (five dimensional
black holes including rotation, and four dimensional black holes without
rotation) 
one can apply the same counting arguments in a quasilocal way 
along the D-brane, to explain the entropy of an inhomogeneous black string.}

Note, however, that this quasilocal picture is lost near the horizon
of the black string.
Even though the horizon always adjusts itself so that its area agrees with
the counting of states, it does so globally - not locally. Locally, the
horizon remains homogeneous, and is largely unaffected by the
traveling waves. This is similar to what happens with other  disturbances
in the spacetime. For example, one can add another sixbrane at a point
$x_0$ outside the black string by simply replacing $H_6$ in (\ref{harm}) 
with
\be 
H_6 = 1 + {r_6\over r} + {g\over 2 |x-x_0|}
\ee
It is clear that the effect of this new sixbrane becomes negligible near
the horizon $r=0$; both the area and local geometry of
the horizon remains unchanged. This is expected
since the area of the horizon is a measure of the number of
internal states, and it should not be affected by the
presence of matter (such as the new sixbrane) with which the
string does not interact.

Let us now return to the six dimensional black string with angular momentum.
Recall that in
section \ref{rot} we imparted angular momentum to
our string by adding the wave $\gamma(u)$, which left
the linear momentum density unchanged.  However, as 
described in \cite{cmp,DGHW}, one may also give
the black string angular momentum by adding appropriate
external waves $h_i(u)$.  Roughly speaking, the wave $\gamma(u)$
corresponds to spinning the string, while the external wave
$h_i(u)$ corresponds to a string gyrating so as to act like a
rotating helical coil.  If we specify both the 
angular and the linear momentum  density of the string, the contribution
to the angular momentum from the `spin waves' $\gamma(u)$
and the `gyrating external waves' $h_i(u)$ are uniquely determined.
We have seen that the Bekenstein-Hawking 
entropy of the resulting spacetime
agrees with the counting of weak coupling bound
states whenever our mesoscopic picture is well-defined,
for any combination of gyration and rotation.

In addition, it is illustrative to consider a different point of
view.  Suppose one wanted to 
study rotating black {\it holes}.  Both 
spinning and gyrating black strings reduce to rotating black holes
in five dimensions, at least after appropriate averaging
(see \cite{cmp,DGHW}).  However, when considering the reduced
solution it does not seem appropriate to specify the
{\it distribution} of momentum along the string.  As a result, we
wish to consider the collection of D-brane
states corresponding to fixed total momentum $P$
and angular momentum $J$, but without any further restriction.
Such states will in general include both
spin waves and gyrating waves; we would like to know which
contribution dominates (if any) and to see that the result
remains compatible with the entropy of the five dimensional black
hole.

Suppose that we consider the somewhat smaller collection of
states for which the spin waves contribute an angular momentum 
$J_{spin}$ and the gyrating waves contribute an angular momentum
$J_{gyro}$, where $J = J_{spin}+ J_{gyro}$.  The relative contributions
of such states
can be determined from the corresponding entropy.  Recall
that both waves effect the entropy by reducing the longitudinal
momentum that may be freely distributed among the various modes.
{}From section \ref{rot}, the spin angular momentum reduces
the available longitudinal momentum by $J_{spin}^2 \kappa^2/Lr_0^4$.
On the other hand, the  gyrating wave reduces  the longitudinal
momentum by term of the form 
$J_{gyro}^2 \kappa^2/Lr_0^2A^2$ which depends on the amplitude $A$
of the wave.  This is to be
expected, as an angular momentum $J$ typically contributes an
energy of the form $J^2/I$ where $I$ is a moment of inertia
($Lr_0^2/\kappa^2$ is the mass of the string).
The division of the angular momentum into spin and gyration
is determined by maximizing the entropy (i.e. the available longitudinal
momentum) subject to the constraint
$J = J_{spin} + J_{gyro}$.

Clearly, the result depends on the allowed radius $A$ of the gyrations.
Let us suppose that the gyrations can be arbitrarily large.  In
this case, the associated black string is quite far from being
translationally invariant and appears to be shaking wildly.
The averaging process that gives the five dimensional spacetime
must then become ill-defined.  It seems unlikely than any observer
could be `effectively five-dimensional' in such a setting; any
observer would find large discrepancies from the predictions of the
five dimensional black hole solution.  As a result, 
such states do not correspond to our picture of a five dimensional
black hole.  The setting of the problem thus restricts
the amplitude $A$ of the gyrations to be much less than $r_0$, the
radius of the five dimensional black hole.  With the condition
$A \ll r_0$, the entropy is maximized
for $J_{spin}=J$, $J_{gyro} = 0$.  This value is then overwhelmingly
likely and the entropy of the entire collection of states with
$J_{spin} + J_{gyro} =J$ and $A \ll r_0$ is
\begin{equation}
S = \sqrt{2 \pi Q_1 Q_5} \sqrt{L  P - J^2 \kappa^2/ r_0^4 }
= 2\pi \sqrt{Q_1 Q_5 N - J^2},
\end{equation}
in agreement with the entropy of the five dimensional black hole
\cite{bmpv}.

\acknowledgements
It is a pleasure to thank D. Lowe  and A. Strominger
for useful discussions. This
work was supported in part by NSF grant PHY95-07065.

\end{document}